\documentclass[nofootinbib, amsmath,amssymb, aps, twocolumn]{revtex4}

\usepackage{hyperref}

\usepackage{graphicx}
\usepackage{dcolumn}
\usepackage{bm}

\usepackage{booktabs}
\usepackage{topcapt}

\usepackage{slashed} 

 \usepackage{axodraw2}

 \usepackage{ulem}
 \usepackage{cancel}
 
 \usepackage{calligra}

\begin{document}
\title{
The soft-gluon limit of the Landau gauge ghost-gluon vertex: results for pure Yang-Mills SU(3) theory from lattice simulations}
\author{Nuno Brito}
\affiliation{Centre for Mathematical Sciences, University of Plymouth, Plymouth, PL4 8AA, United Kingdom}
\author{Orlando Oliveira}
\author{Paulo J. Silva}
\affiliation{CFisUC, Departament of Physics, University of Coimbra, 3004-516 Coimbra, Portugal}

\begin{abstract}

This work reports on the computation of the ghost-gluon vertex for the pure Yang-Mills SU(3) gauge group,
using lattice simulations and the Landau gauge. 
The simulations access only one of the form factors that describes the vertex.
The form factor is estimated using large statistical ensembles of gauge configurations, 
with two different lattice spacings and two different volumes to check for finite size effects. 
Moreover, the calculation shows the importance of using lattice perturbation theory, instead of 
its continuum version, to correct for the breaking of rotational symmetry. 
The measured bare lattice form factors are compatible, within errors, for all the ensembles.
The form factor has a maximum at momentum $\sim 1$ GeV, is suppressed in the infrared and is
compatible with a constant behaviour at high momenta, in good agreement with the corresponding lattice estimations for the SU(2) gauge group.

\end{abstract}

\maketitle
\tableofcontents

\maketitle

\section{Introduction and Motivation}

The modern understanding of the phenomenology of elementary particles, 
from the observations in high energy physics colliders to nuclear physics, 
requires the solution of Quantum Chromodynamics (QCD) for various regimes.
Its low energy regime, associated with e.g. bound states, calls for non-perturbative solutions of the theory, 
while its high energy regime relies on perturbative methods.
An example coming from high energy phenomenology associated with the non-perturbative regime
is the recent claim of a toponium like state observed by the ATLAS and CMS experiments at LHC, see~\cite{Pintucci:2026agl} 
and references therein. 
Atomic nuclei are states whose dominant components are many-body quark and gluon states bound by 
the strong interaction. 
These states are also relevant to investigate the quark-gluon plasma, that is expected to occur
at high temperatures and densities. 
Despite the enormous progress of the past years, the study of the non-perturbative regime of 
Quantum Chromodynamics is still a challenge for theoretical physics. 

From the theory side, the investigation of the low energy regime of QCD relies either on its continuum formulation,
that uses the Dyson-Schwinger equations (DSE), other integral equations or the renormalization group approach, 
or on the formulation of the theory on a finite set of points, i.e. on lattice QCD simulations. 
Another way out to understand hadronic matter is to model the strong interaction via an effective field theory 
tuned for the low energy regime.
It is the interplay of all the approaches, by combining the results from the various techniques or by matching each 
other, that has proved to be a good guide to understand the theory and the experimental results.
 As a starting point, it is common to consider the two point correlation functions in QCD, 
namely the gluon, the ghost and the quark propagators.

In recent years, the infrared properties of Landau-gauge QCD Green's functions have motivated numerous studies, 
employing both lattice 
simulations~\cite{Cucchieri:2007rg,Cucchieri:2008fc,Bogolubsky:2009dc,Duarte:2016iko,Dudal:2018cli} and continuum 
approaches such as DSE~\cite{Alkofer:2000wg,Fischer:2003zc,Fischer:2008uz,Aguilar:2008xm,Binosi:2009qm,Boucaud:2008gn,Boucaud:2008ji,Fischer:2009tn,Alkofer:2008dt,Zierler:2023qvz}, 
the refined Gribov-Zwanziger formalism~\cite{Dudal:2005na,Dudal:2007cw,Dudal:2008rm,Dudal:2010fq}, 
and effective models like the Curci-Ferrari 
model~\cite{Tissier:2010ts,Tissier:2011ey,Pennington:2011wd}.
Among the fundamental two-point correlation functions, the ghost and gluon propagators have been extensively studied 
and are now relatively well understood from both lattice and continuum perspectives. 

A complete non-perturbative description also requires a precise knowledge of the three-point correlation functions. 
In QCD the three-point functions are the three-gluon, the quark-gluon and the ghost-gluon vertices. 
The quark-gluon vertex has a leading role in the quark dynamics. 
On the other hand, the three-gluon vertex is at the heart of our understanding of asymptotic freedom and 
of dynamical mass generation \cite{Ferreira:2025anh}. 
The ghost-gluon vertex is fundamental to set the infrared behaviour 
of Yang-Mills Green's functions~\cite{Boucaud:2011eh,Dudal:2012zx}. As an example, the vertex is used as an input in \cite{Huber:2025kwy} where the glueball masses in SU(3) Yang-Mills are obtained from functional approaches (DSE and Bethe-Salpeter equations).  

This work addresses the lattice computation of the ghost-gluon vertex for pure Yang-Mills SU(3) in the Landau gauge.
Hopefully, the results reported herein will contribute to a deeper understanding of QCD, and provide guidance for more accurate modelling of the ghost–gluon vertex in continuum approaches \cite{Boucaud:2009ga, Boucaud:2006xe, Sternbeck:2009yx}.
Recall that Taylor’s non-renormalization theorem for the Landau gauge 
ghost-gluon vertex~\cite{Taylor:1971ff} motivates the use of a constant vertex within the Dyson-Schwinger continuum approach. 
The kinematics of Taylor's theorem is not reachable in lattice simulations, and it can not be tested 
directly with lattice techniques. However, having precise lattice data over a range of momenta
can impact the numerical solutions of the Dyson-Schwinger equations.

The lattice studies for the SU(2) gauge group, that consider various kinematical configurations, 
report a vertex that deviates from the bare vertex~\cite{Cucchieri:2004sq,Cucchieri:2006tf,Cucchieri:2008qm}, and 
that is suppressed at low momenta. 
In the continuum formulation, the Refined Gribov-Zwanziger effective action~\cite{Mintz:2017qri,Barrios:2024idr} 
and the Curci-Ferrari model~\cite{Barrios:2020ubx,Figueroa:2021sjm}
also predict an infrared suppression for the gluon-ghost vertex for SU(2) and SU(3) gauge theories. 
Moreover, the analysis of the Dyson-Schwinger equation for the SU(3) ghost propagator using lattice inputs for the 
propagators, solved for the ghost-gluon vertex, also shows infrared suppression of this 
fundamental vertex~\cite{Dudal:2012zx}.

While the available bare lattice data for SU(2) cover various kinematical configurations with good statistical precision, the same does not hold for SU(3). More precisely, in~\cite{Ilgenfritz:2006he, Sternbeck:2006rd} the ghost-gluon vertex was also studied using SU(3) lattice simulations in the Landau gauge.
However, the quantities computed therein refer to a particular renormalization scheme and the bare data is not reported. 
Indeed, the available SU(3) lattice data do not provide much information on the 
form factors associated with the ghost-gluon vertex. Herein, we aim to fill this gap in the literature and
provide bare lattice data for one of the ghost-gluon vertex form factors; see below for discussions.
Furthermore, our study looks also at the finite volume effects, 
and at the effects associated with the breaking of rotational symmetry.

Our main goal is to  deliver accurate information on the ghost-gluon vertex 
for the pure gauge Yang-Mills theory with the SU(3) gauge group. 
In this first manuscript, only the soft gluon limit, i.e. the case where the gluon momentum vanishes, will be considered. 
The computation of other kinematics is planned and will be reported in future publications.
The calculation reported herein relies partially on the techniques used in the computation of the lattice
Batalin-Vilkovisky function~\cite{Aguilar:2024bwp}. The gauge ensembles considered are a subset of the 
configurations generated in the investigation of the four-gluon vertex~\cite{Colaco:2024gmt,Oliveira:2025tfq}
that have also been used in~\cite{Oliveira:2025rzw} to investigate the gluon propagator and finite size effects.
Preliminary results for the ghost-gluon vertex can be found in~\cite{Brito:2024aod,Brito:2025kfz}.

The manuscript is organized as follows. In Sec. \ref{Sec:Def} the Green functions to be considered here are defined and the extraction of the relevant form factor is described in the continuum. In Sec. \ref{Sec:Lat} the lattice setup and techniques exployed to extract the desired form factors are presented. In Sec. \ref{Sec:Results} we present and discuss the results obtained. In Sec. \ref{Sec:Conclusion} we report on the key takeaways from this work.

\section{Continuum Definitions} 
\label{Sec:Def}

\begin{figure}[t]
\centering
\includegraphics[width=3.5in]{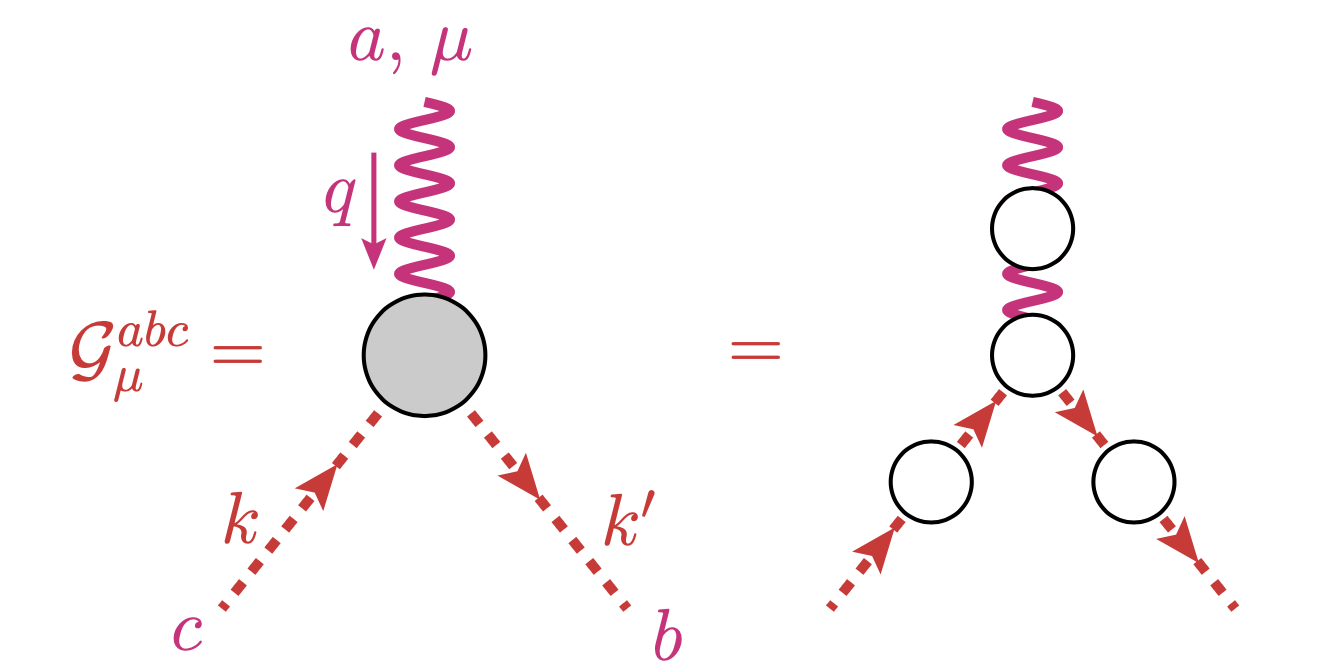}
\caption{The full ghost-gluon Green function $\mathcal{G}^{abc}_\mu$ in momentum space (left) 
               and its decomposition in terms of one-particle irreducible diagrams (right) represented by empty blobs.
              Feynman diagrams were built using the Axodraw package \cite{Collins:2016aya}.}
\label{Fig:GreenFunction}
\end{figure}

The primary quantity computed with lattice QCD simulations is the full Green function
\begin{equation}
\label{eq:full_greens_function}
\mathcal{G}^{abc}_\mu(x, \, y, \, z) ~ = ~\langle A^a_\mu (x) ~ c^b(y) ~\overline c^c(z) \rangle \ ,
\end{equation}
where $\langle \cdots \rangle$ stands for vacuum expectation value, that is realized as an ensemble average.
The diagrammatic representation of this Green function in the momentum space is given in Fig.
\ref{Fig:GreenFunction}, together with its decomposition in terms of one-particle irreducible functions, that
are represented as empty blobs. In momentum space, the full Green function is written as
\begin{eqnarray}
\mathcal{G}^{abc}_\mu (k, \, k^\prime ; \, q) & = & D^{aa^\prime}_{\mu\mu^\prime}(q) ~G^{b b^\prime} (k^\prime) ~ G^{c c^\prime} (k) 
 \nonumber \\
 & & \qquad 
 ~ \Gamma^{a^\prime b^\prime c^\prime}_{\mu^\prime}(k, \, k^\prime ; \, q)
\end{eqnarray}
where 
\begin{eqnarray}
  D^{ab}_{\mu\nu}(q) = \delta^{ab} \, D_{\mu\nu}(q) \quad\mbox{ and }\quad G^{ab}(q) = \delta^{ab} \, G(q^2) , 
  \label{propagadores}
\end{eqnarray}
are the gluon and the ghost propagators, respectively. The one-particle irreducible Green function associated with the ghost-gluon vertex
can be written in terms of two Lorentz scalar form factors $H_1(k^2, \, {k^\prime}^2 ; \, q^2)$ and $H_2(k^2, \, {k^\prime}^2 ; \, q^2)$ as
\begin{equation}
 \Gamma^{a b c}_{\mu}(k, \, k^\prime ; \, q) ~ = ~   \,
 g \, f_{abc} \, \Big( k^\prime_\mu \, H_1  -  q_\mu \, H_2 \Big).
\end{equation}
In the Landau gauge the gluon propagator is orthogonal and, therefore, 
\begin{eqnarray}
\mathcal{G}^{abc}_\mu (k, \, k^\prime ; \, q) & = & g \, f_{a^\prime b^\prime c^\prime} ~ G^{b^\prime b} (k^\prime) ~ G^{c^\prime c} (k) 
 \nonumber \\
 & & \quad 
  \Big( k^\prime_{\mu^\prime}  D^{aa^\prime}_{\mu\mu^\prime}(q) \Big)  ~ H_1(k^2, \, {k^\prime}^2 ; \, q^2) 
  \nonumber \\
  & = &
   g \, f_{a b c} ~ G({k^\prime}^2) ~ G(k^2) 
 \nonumber \\
 & & \quad 
  \Big( k^\prime_{\mu^\prime}  D_{\mu\mu^\prime}(q) \Big)  ~ H_1(k^2, \, {k^\prime}^2 ; \, q^2) 
  \ .
\end{eqnarray}
Since the one-particle irreducible Green function at tree-level reads
\begin{equation}
 \left( \Gamma^{(0)}\right)^{a b c}_{\mu}(k, \, k^\prime ; \, q) ~ = ~  
 g ~ f_{abc}  ~ k^\prime_\mu \ ,
 \label{Gamma:Cont_TL}
\end{equation}
it follows that
\begin{eqnarray}
\label{eq:h1_contraction_cont}
& & 
H_1(k^2, \, {k^\prime}^2 ; \, q^2)  ~ = ~
\nonumber \\
& &
\qquad
 = \frac{ \Gamma^{(0)} \cdot \mathcal{G}}{\Gamma^{(0)} \cdot \Gamma^{(0)} } 
~
 \frac{1}{ \left( k^\prime_\mu k^\prime_\nu  D_{\mu\nu}(q) \right) ~ G (k^\prime) ~ G (k) }
  \ .
   \label{H1:Cont_TL}
\end{eqnarray}
In lattice perturbation theory the tree-level expression for $\Gamma^{(0)}$ differs from that given in
Eq. (\ref{Gamma:Cont_TL}) and the computation of $H_1$ requires minimal modifications
of Eq. (\ref{H1:Cont_TL}); see Sec. \ref{Sec:Lat} for discussions.

\section{Lattice Definitions and Setup}
\label{Sec:Lat}

The simulations reported here consider the pure Yang-Mills theory and the Wilson gauge action for the SU(3) gauge group.
They were performed with Chroma~\cite{Chroma} and PFFT~\cite{PFFT} libraries. 
For gauge fixing,
a Fourier accelerated steepest descent method inspired in~\cite{Davies:1987vs} was used,
with a stopping criterion defined by the average value of the lattice version of the Landau gauge condition,
that should be smaller than $10^{-15}$ in lattice units. 
Further details on definitions and on the gauge fixing algorithm can be found in~\cite{Silva:2004bv}.
The statistical errors were computed with the bootstrap method setting the confidence level to be 67.5\%.

The computations of the vertex use 5000 gauge configurations for each of the following ensembles:
\begin{eqnarray}
&&
\beta = 6.0, \hspace{1cm} 32^4, \hspace{1cm} 48^4 , \nonumber \\
&&
\beta = 6.2, \hspace{1cm} 48^4, \hspace{1cm} 64^4 .
\end{eqnarray}
For the conversion into physical units we use  $1/a(\beta = 6.0) = 1.943$ GeV ($a (\beta = 6.0) = 0.1016$ fm) 
and $1/a(\beta = 6.2) = 2.7052$ 
($a (\beta = 6.2) = 0.0729$ fm) that, as described in~\cite{Oliveira:2025rzw}, were set to provide the better 
agreement between the lattice gluon propagator 
computed for the two $\beta$ values. Then, the length of each lattice is
\begin{eqnarray}
&&
\beta = 6.0, \hspace{1cm} 3.25 ~\mbox{fm}, \hspace{1cm} 4.87 ~\mbox{fm}, \nonumber \\
&&
\beta = 6.2, \hspace{1cm} 3.50 ~\mbox{fm}, \hspace{1cm} 4.67 ~\mbox{fm},
\end{eqnarray}
respectively. The four sets correspond to two different physical volumes and two different lattice spacings  and, therefore, allow to check for lattice and 
volume effects on the form factor.

The computation of the $H_1$ requires the evaluation of two-point correlations functions.
In the Landau gauge, the lattice gluon propagator is defined as
\begin{equation}
\langle A^a_\mu (q^\prime) A^b_\nu (q) \rangle = V ~ \delta ( q^\prime + q )  ~ D^{ab}_{\mu\nu}(q)
\end{equation}
and is orthogonal to the gluon momentum, i.e. $q_\mu D^{ab}_{\mu\nu}(q) = q_\nu D^{ab}_{\mu\nu}(q) = 0$.
Its tensorial structure is given in Eq. (\ref{propagadores}), with
\begin{equation}
D_{\mu\nu} (q) = \left( \delta_{\mu\nu} - \frac{q_\mu q_\nu}{q^2} \right) ~ D(q^2)
\end{equation}
In a lattice simulation, the breaking of rotational symmetry replaces $D(q^2)$
by a function of $q^2$ and other lattice 
invariants~\cite{Leinweber:1998im,Leinweber:1998uu,Becirevic:1999uc,deSoto:2007ht,Vujinovic:2018nqc,Catumba:2021hcx,Oliveira:2025rzw} and the same considerations apply to any lattice Green function.
The lattice propagator, which we write as $\widetilde D(q^2, \dots)$ with the dots 
standing for the remaining dependences, reduces to $D(q^2)$ in the continuum 
limit. In practice, it is difficult to perform the continuum limit of the propagator, see e.g.~\cite{Oliveira:2012eh}, 
and it is common to represent the lattice propagator only for certain classes of momenta.
The momentum cuts used are two fold. To suppress the effects due to the breaking of rotational symmetry 
in the UV regime, above $0.7$ GeV, only those momenta that are selected by the cylindrical and conical cuts~\cite{Leinweber:1998uu} are taken into account. On the other hand, to have a good covering of the low momenta region~\cite{Dudal:2018cli}, 
all lattice data points for momenta below $ 0.7$ GeV are considered.

In the computation of the propagator, we take for the gluon field 
\begin{equation}
A_\mu ( x + a \, \hat{\mu}/2) ~ = ~
\left. \frac{U_\mu (x) - U^\dagger_\mu(x)}{2 \, i \, a \, g} \right|_{\mbox{traceless}}
\end{equation}
here $a$ is the lattice spacing, $g$ the bare lattice coupling constant, $\hat{\mu}$  the unit vector along the lattice direction $\mu$, and $U_\mu$
are the gauge link. In momentum space, the gluon field is given by
\begin{equation}
A_\mu (q) ~ = ~ \sum_x ~e^{-i q (x +  a \, \hat{\mu}/2)} ~A_\mu ( x + a \, \hat{\mu}/2) 
\end{equation}
where the naive lattice momentum reads
\begin{equation}
q_\mu ~ = ~  \frac{ 2 \, \pi }{a \,L_\mu} n_\mu ,
\qquad 
n_\mu ~ = - L_\mu / 2 + 1, \, \dots , \, L_\mu / 2
\label{ImpMom}
\end{equation}
where  $L_\mu$ is the number of lattice points on direction $\mu$.
The improved momentum 
\begin{equation}
p_\mu ~ = ~ \frac{2}{a}  \, \sin \left( \frac{\pi}{ L_\mu } \, n_\mu \right)  = q_\mu + \mathcal{O}(a^2)
\end{equation}
is commonly used in the study of the lattice propagator as it helps reducing the effects due to the breaking of rotational symmetry,
improves the orthogonality of the propagator and is also the natural momenta that appears in lattice perturbation
theory --- see e.g ~\cite{Catumba:2021hcx} and references therein for discussions and further details. Herein, we consider
only the zero momentum propagator, whose tensor structure is simpler and reads
\begin{equation}
D_{\mu\nu} (0) ~ = ~ \delta_{\mu\nu} D(0).
\end{equation}

The ghost propagator is defined as the inverse of a suitable discretization of the continuum Faddev-Popov operator $M$, see e.g.~\cite{Suman:1995zg,Sternbeck:2006rd}, 
which in momentum space is given by
\begin{equation}
\label{eq:ghost_propagator}
G^{ab}(q) ~ = ~  \int d^4 x ~ e^{ -i \, q \, (x -y) } ~ \langle 0 | M^{-1} (x,y) |0 \rangle \ .
\end{equation}
The lattice Faddeev-Popov operator $M^{ab}(x,y)$ is singular but can be inverted in the subspace orthogonal to its zero modes. 
The inverse problem in the orthogonal subspace can be transformed into a sparse linear system of equations that can be solved 
with the conjugate gradient method. The ghost propagator was computed with the method described 
in~\cite{Suman:1995zg}, using a single source, at the origin of the lattice, when solving the associated linear system.

To access the form factors associated with the one-particle irreducible ghost-gluon Green function, 
the full Green's functions, see Eq.~(\ref{eq:full_greens_function}), has to be contracted as described
in Eq.~(\ref{H1:Cont_TL}). The following relation
\begin{eqnarray}
    \label{eq:ko_point_source}  
     \sum_{c} f_{abc}A^c_\mu(x) = && -\frac{1}{2g}\operatorname{Tr}\left(\left[t^a,t^b\right]\right.
    \\
    && \nonumber \left.\left\{\left(U_{-\mu}^\dagger(x) + U_{\mu}(x)\right) - h.c\right\}\right)
\end{eqnarray}
helps in doing the contractions.
In Eq.~(\ref{H1:Cont_TL}), the numerator and each of the propagators are given by ensemble averages.

In lattice perturbation theory, the tree-level lattice ghost-gluon one-particle irreducible Green function reads
\begin{equation}
 \left( \Gamma^{(0)}_{\textrm{Lat}}\right)^{a b c}_{\mu}(k, \, k^\prime ; \, q) ~ = ~  
 g ~ f_{abc}  ~ \widetilde{k}^\prime_\mu ~\cos(k^\prime_\mu a /2) \ ,
 \label{Gamma:Lat_TL}
\end{equation}
where
\begin{equation}
 \widetilde{k}^\prime_\mu ~= ~ \frac{2}{a} \, \sin \frac{ a \, k^\prime_\mu}{2} 
 \quad\mbox{and}\quad  k^\prime_\mu ~=~ \frac{2 \, \pi}{a \, L_\mu} \, n_\mu
\end{equation}
with $n_\mu$ is as in Eq. (\ref{ImpMom}). In the computation of $H_1$, see Eq. (\ref{H1:Cont_TL}), one can use either
the continuum tree-level vertex (\ref{Gamma:Cont_TL}) or its lattice version (\ref{Gamma:Lat_TL}). In the latter case,
in the denominator of  (\ref{H1:Cont_TL}) the quantity $k^\prime_\mu k^\prime_\nu  D_{\mu\nu}(q)$ is replaced by
\begin{equation}
 \sum_{\mu \, \nu} ~ \widetilde{k}^\prime_\mu \, \cos(k^\prime_\mu a /2) ~ \widetilde{k}^\prime_\nu  \, \cos(k^\prime_\nu a /2)D_{\mu\nu}(q) \ .
 \end{equation}

In the following, the ghost–gluon form factor obtained using Eq. (\ref{Gamma:Cont_TL}) will be denoted by  $H_1^{(\mbox{c})}$ , while Eq. (\ref{Gamma:Lat_TL}) will be used to define $H_1^{(\mbox{l})}$.
Moreover, in all cases, only the improved momentum will be used to label $H_1$.

\section{Results for the Vertex}
\label{Sec:Results}

\begin{figure}[t] 
   \centering
   \includegraphics[width=3.5in]{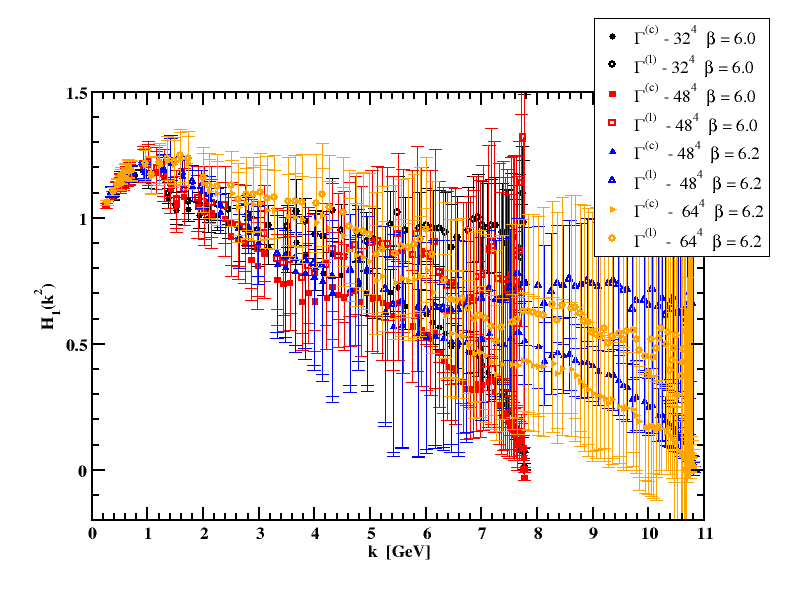} 
   \caption{Bare $H_1(k^2)$ for all ensembles and volumes. The full symbols were computed using $\Gamma^{(c)}$ to project the form factor, while the open symbols
               use $\Gamma^{(l)}$ in the evaluation of $H_1$.}
   \label{fig:H1All}
\end{figure}

\begin{figure*}[t] 
   \centering
   \includegraphics[width=6in]{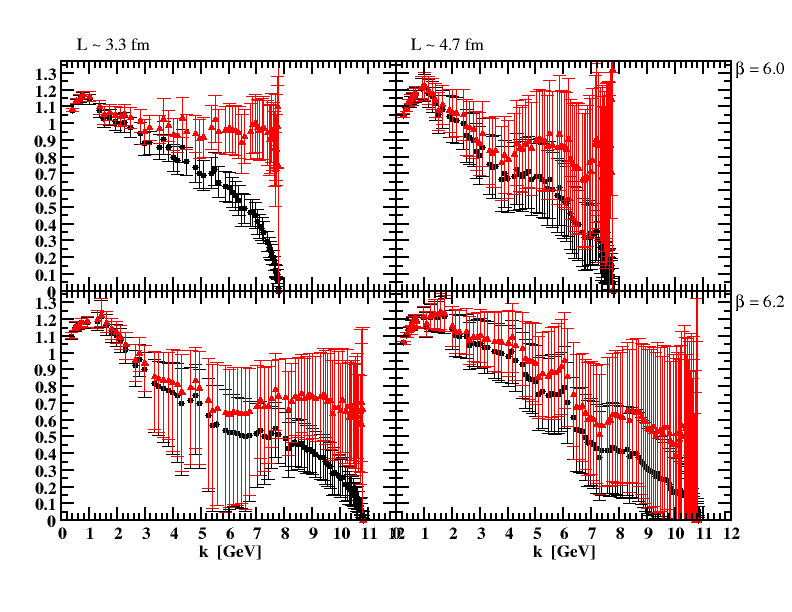} 
   \caption{Bare $H_1(k^2)$ for all  ensembles and volumes. The results associated with $\Gamma^{(c)}$ are the full black circes, while those associated with
                 $\Gamma^{(l)}$ are the full red triangles.}
   \label{fig:H1All4x4}
\end{figure*}

\begin{figure}[t] 
   \centering
   \includegraphics[width=3.5in]{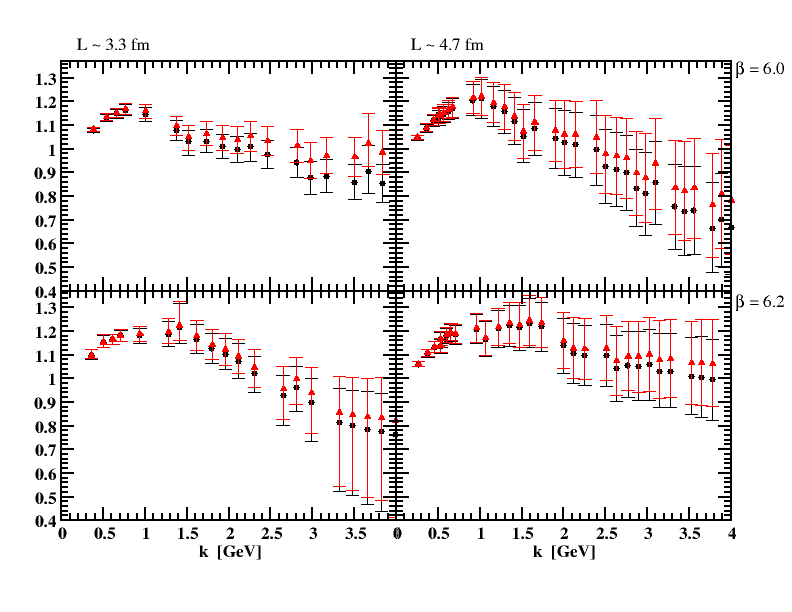} 
   \caption{Bare $H_1(k^2)$ for all  ensembles and volumes and up to 4 GeV.}
   \label{fig:H1All4x4-up4GeV}
\end{figure}

The bare $H_1(k^2)$ form factor for all ensembles and for data sets that comply with the momentum cuts defined above is 
reported in Figs \ref{fig:H1All}, \ref{fig:H1All4x4} and \ref{fig:H1All4x4-up4GeV}.
As shown, there are significant differences when using either $\Gamma^{(c)}$ or $\Gamma^{(l)}$ to extract the form 
factor from the full Green function. The major difference takes place at larger momenta,
where $\Gamma^{(c)}$ undestimates $H_1$, relative to the outcome of using $\Gamma^{(l)}$.
If from $\Gamma^{(c)}$ the prediction is a decreasing $H_1(k^2)$, the use of $\Gamma^{(l)}$ 
returns a form factor that, in all cases, is compatible with a constant value. 
It is only when $\Gamma^{(l)}$ is used to measure $H_1$ that the lattice data becomes
compatible with the predictions of continuum perturbation theory at large momentum. At the smallest momenta 
(the lattice data starts at momentum $\sim 265$ MeV), there is no difference between $H_1$ measured from using 
either of the vertices. We take these results as an indication that lattice effects are important for large momenta and
that $\Gamma^{(l)}$ should always be used to minimize the effects of the breaking of rotational symmetry.

The observed pattern in the statistical errors can be understood from the way that $H_1$ is evaluated, see 
Eq. (\ref{H1:Cont_TL}). Indeed, in the soft gluon limit the full Green function has to be divided by $D(0) \, G^2(k^2)$.
Given that $G(k^2)$ decreases as $k^2$ is increased, by dividing by smaller numbers it is expected an increase 
in the statistical errors for larger momenta. 
This reproduces the observed pattern and is particularly clear in the output of the simulations of the smaller physical
volume with $\beta = 6.0$. Recall that for the same $\beta$ value, larger volumes imply larger fluctuations. 
 Note that in the computation of ghost propagator only a single point source was used, and a combination of the 
 outcome of using various sources can significantly improve the signal-to-noise ratio in the evaluation of $H_1$. 
 Note that, if one goes beyond the soft gluon limit, one has to take into account also the fluctuations associated with
 the gluon propagator function $D(q^2)$. They will decrease the signal-to-noise ratio at larger momenta as
 the values for $1 / D(q^2) \, G(k) \, G(k^\prime)$ will decrease faster at higher momenta.
For example, this type of behavior was observed in the lattice simulation where the four-gluon one-particle irreducible 
form factors were computed~\cite{Colaco:2024gmt,Oliveira:2025tfq}.

Despite the differences already mentioned due to the use of $\Gamma^{(l)}$ versus $\Gamma^{(c)}$, looking only
at the results associated with $\Gamma^{(l)}$, according to Figs \ref{fig:H1All},\ref{fig:H1All4x4} and \ref{fig:H1All4x4-up4GeV}, 
the various estimations of $H_1$ are compatible for all the simulations considered. 
We take this as an indication that finite size and volume effects are small or that they are associated with effects
that are beyond the statistical resolution of the simulations.

The form factor $H_1(k^2)$  is compatible with a constant value for large momenta, 
decreases when the  zero momentum is approached, 
and has a maximum at momentum $k \sim 1$ GeV.
This maximum corresponds to an
enhancement of about 20\% relative to the infrared and ultraviolet values.
 Despite the limited access for momenta below 1 GeV, the data provides a clear answer on the deep infrared  behavior of $H_1(k^2)$, where the form factor is suppressed.
The observed $H_1(k^2)$ is in good agreement with the results computed for the SU(2) gauge 
group~\cite{Cucchieri:2004sq,Cucchieri:2006tf,Cucchieri:2008qm}. Futhermore, in \cite{Zierler:2023qvz} the authors obtain $H_{1}$ from a set of truncated DSE equations which displays the same functional behaviour as the results in this work.

\subsection{Modelling the ghost-gluon vertex}

The form factor $H_1$ was estimated from the analysis of the ghost Dyson-Schwinger equation in~\cite{Dudal:2012zx} 
or from an one-loop dressed approximation in~\cite{Aguilar:2010cn,Rojas:2013tza,Oliveira:2020yac}. 
Although, in these and subsequent works it was assumed that $H_1$ depends not in ghost momentum
but in the gluon momentum, the gross features of the form factors are similar to those observed here. 
Indeed, they predict that $H_1$ has a maximum at the gluon momentum $\sim 1$ GeV, is constant in the 
ultraviolet regime and decreases from the maximum for gluon momentum smaller than $\sim 1$ GeV 
and when it approaches the zero momentum. 
Despite the dependence of the form factor being on the gluon momentum and not on the ghost momentum, given the 
observed qualitative results and following~\cite{Dudal:2012zx}, it seems reasonable to try to describe our lattice data
with the same type of functional form
\begin{equation}
H_1(k^2) = Z \Bigg(  c \left( 1 + \frac{a^2 \, k^2}{k^4 + b^4}\right)  + (1 - c) \frac{w^4}{w^4 + k^4}  \Bigg) \ ,
\end{equation}
where $Z$ is an overall normalization factor. It turns out that this functional form is able to reproduce well the lattice data. However, the fitted parameters are not always
compatible within the different ensembles
and in the second term $w$ is sometimes compatible with zero. Therefore, to summarize the results we provide instead the outcome of the fits with only the first term:
\begin{table}[h]
   \centering
   \begin{tabular}{r@{\hspace{0.25cm}}r@{\hspace{0.5cm}}r@{\hspace{0.5cm}}r@{\hspace{0.5cm}}l@{\hspace{0.25cm}}l@{\hspace{0.25cm}}l}
 $\beta$ & L & d.o.f & $\chi^2_{\rm red}$ & $c$ & $a^2$ & $b^4$ \\
    6.0 & 32 &   51 &  0.19 &   0.9786(65) &   0.215(16) &  0.289(26) \\
    6.0 & 48 &   76 &  0.33 &   0.9902(85) &   0.243(34) &  0.314(59) \\
    6.2 & 48 &   74 &  0.89 &   0.888(21)  &    0.599(93) &  0.553(95) \\
    6.2 & 64 &   98 &  0.71 &   1.009(11)  &    0.365(66) &   0.59(14) 
   \end{tabular}
\end{table}

\noindent
where the third column is the number of degrees of freedom, the fourth is the $\chi^2_{\rm red} = \chi^2/d.o.f.$ and the dimensionfull quantities $a^2$ and $b^4$ are given in powers of GeV.

\section{Summary and Conclusions}
\label{Sec:Conclusion}

  In this work a lattice calculation of the Landau gauge ghost-gluon vertex using quenched SU(3) gauge configurations in the soft-gluon limit is discussed. The results reported consider four different large statistical ensembles of configurations
  that differ in their physical volume or in the value of the lattice spacing. 
  No significant difference between the results coming from the various simulations is observed, suggesting that
  the finite size effects for the volumes and lattice considered are small or negligible.

  The form factor associated with the ghost-gluon vertex shows essentially the same features already seen in 
   pure Yang-Mills SU(2) lattice simulations. In particular, the lattice data is compatible with a constant value 
   in the UV region, shows a maximum around 1 GeV and decreases towards has the infrared limit is approached. 
   Its UV behaviour, despite the relative large statistical errors, is compatible with a constant value,
   a feature is become clear only when the form factor measure uses the tree-level expression that comes from 
   lattice perturbation theory. This result highlights the relevance of lattice corrections.

   The statistical precision achieved allow to investigate the compatibility of the lattice data with a 
   functional form proposed in~\cite{Dudal:2012zx}. Recall that this functional form was associated with
   the dependence on the gluon momenta and not on the ghost. 
   As described, the functional form reproduces well the lattice data, 
   in the sense that the values $\chi^2/d.o.f. \sim 1$. 
   This result should be read with a grain of salt, as the fitting parameters are not compatible within 
   one standard deviation when the figures for the various ensembles are compared.
    However, note that the parameter values are in the same ballpark. 
   The differences can probably be resolved by extrapolating to the infinite volume, an attempt not pursued given the limited number of lattice simulations.
   
   We plan to extend this study beyond the soft-gluon limit, not only by exploring
   other momentum configurations, but also to go beyond the Landau gauge.

\section*{Acknowledgments}

This work was financed through national funds by FCT - Fundação para a Ciência e Tecnologia, I.P. in the framework of the projects UIDB/04564/2020 , UIDP/04564/2020 and UID/04564/2025, with DOI identifiers  \url{10.54499/UIDB/04564/2020}, \url{10.54499/UIDP/04564/2020} and \url{10.54499/UID/04564/2025} . 
P. J. Silva  acknowledges financial support from FCT contract CEECIND/00488/2017, with DOI identifier \url{10.54499/CEECIND/00488/2017/CP1460/CT0030}.
NB is supported by the Science and Technology Facilities Council (STFC) Consolidated Grant No. ST/X508676/1. 

The authors acknowledge the Laboratory for Advanced
Computing at the University of Coimbra (\url{http://www.uc.pt/lca}) and the Minho Advanced
Computing Center (\url{http://macc.fccn.pt}) for providing access to the HPC resources. 
Access to Navigator was partly supported by the FCT Advanced Computing Projects 2021.09759.CPCA, 2022.15892.CPCA.A2 and 2023.10947.CPCA.A2 with DOI identifiers \url{10.54499/2021.09759.CPCA}, \url{10.54499/2022.15892.CPCA.A2} and  \url{10.54499/2023.10947.CPCA.A2} respectively. 
Access to Bob was supported by the FCT Advanced Computing Project CPCA/A2/6816/2020.
Access to Deucalion was supported by the
FCT Advanced Computing Project 2024.11063.CPCA.A3.


\begin{thebibliography}{99}

\bibitem{Pintucci:2026agl}
L.~Pintucci [CMS and ATLAS],
Nuovo Cim. C \textbf{49}, no.1-2, 33 (2026)
doi:10.1393/ncc/i2026-26033-4


\bibitem{Cucchieri:2007rg}
A. Cucchieri and T. Mendes,
\textit{Phys. Rev. Lett.} \textbf{100} (2008) 241601
[arXiv:0712.3517 [hep-lat]].

\bibitem{Cucchieri:2008fc}
A. Cucchieri and T. Mendes,
\textit{Phys. Rev. D} \textbf{78} (2008) 094503
[arXiv:0804.2371 [hep-lat]].

\bibitem{Bogolubsky:2009dc}
I. Bogolubsky, E. Ilgenfritz, M. Muller-Preussker and A. Sternbeck,
\textit{Phys. Lett. B} \textbf{676} (2009) 69
[arXiv:0901.0736 [hep-lat]].

\bibitem{Duarte:2016iko}
A.~G.~Duarte, O.~Oliveira and P.~J.~Silva,
Phys. Rev. D \textbf{94}, no.1, 014502 (2016)
doi:10.1103/PhysRevD.94.014502
[arXiv:1605.00594 [hep-lat]].

\bibitem{Dudal:2018cli}
D.~Dudal, O.~Oliveira and P.~J.~Silva,
Annals Phys. \textbf{397}, 351-364 (2018)
doi:10.1016/j.aop.2018.08.019
[arXiv:1803.02281 [hep-lat]].


\bibitem{Alkofer:2000wg}
R. Alkofer and L. von Smekal,
\textit{Phys. Rept.} \textbf{353} (2001) 281
[hep-ph/0007355].

\bibitem{Fischer:2003zc}
C. S. Fischer and R. Alkofer,
\textit{Phys. Rev. D} \textbf{67} (2003) 094020
[hep-ph/0301094].

\bibitem{Fischer:2008uz}
C. S. Fischer, A. Maas and J. M. Pawlowski,
\textit{Ann. Phys.} \textbf{324} (2009) 2408
[arXiv:0810.1987 [hep-ph]].

\bibitem{Aguilar:2008xm}
A. Aguilar, D. Binosi and J. Papavassiliou,
\textit{Phys. Rev. D} \textbf{78} (2008) 025010
[arXiv:0802.1870 [hep-ph]].

\bibitem{Binosi:2009qm}
D. Binosi and J. Papavassiliou,
\textit{Phys. Rept.} \textbf{479} (2009) 1
[arXiv:0909.2536 [hep-ph]].

\bibitem{Boucaud:2008gn}
P. Boucaud, J. Leroy, A. Le Yaouanc, J. Micheli, O. Pene and J. Rodriguez-Quintero,
\textit{JHEP} \textbf{06} (2008) 099
[arXiv:0803.2161 [hep-ph]].

\bibitem{Boucaud:2008ji}
P. Boucaud, J. Leroy, A. Le Yaouanc, J. Micheli, O. Pene and J. Rodriguez-Quintero,
\textit{JHEP} \textbf{06} (2008) 012
[arXiv:0801.2721 [hep-ph]].

\bibitem{Fischer:2009tn}
C. S. Fischer and J. M. Pawlowski,
\textit{Phys. Rev. D} \textbf{80} (2009) 025023
[arXiv:0903.2193 [hep-th]].

\bibitem{Alkofer:2008dt}
R. Alkofer, M. Q. Huber and K. Schwenzer,
\textit{Phys. Rev. D} \textbf{81} (2010) 105010
[arXiv:0801.2762 [hep-th]].

\bibitem{Zierler:2023qvz}
F.~Zierler and R.~Alkofer,
Phys. Rev. D \textbf{109} (2024) no.7, 074024
doi:10.1103/PhysRevD.109.074024
[arXiv:2312.06463 [hep-ph]].

\bibitem{Dudal:2005na}
D. Dudal, S. Sorella, N. Vandersickel and H. Verschelde,
\textit{Phys. Rev. D} \textbf{72} (2005) 014016
[hep-th/0502183].

\bibitem{Dudal:2007cw}
D. Dudal, J. Gracey, S. Sorella, N. Vandersickel and H. Verschelde,
\textit{Phys. Rev. D} \textbf{77} (2008) 071501
[arXiv:0711.4496 [hep-th]].

\bibitem{Dudal:2008rm}
D. Dudal, J. Gracey, S. Sorella, N. Vandersickel and H. Verschelde,
\textit{Phys. Rev. D} \textbf{78} (2008) 065047
[arXiv:0806.4348 [hep-th]].

\bibitem{Dudal:2010fq}
D. Dudal, O. Oliveira and N. Vandersickel,
\textit{Phys. Rev. D} \textbf{81} (2010) 074505
[arXiv:1002.2374 [hep-lat]].

\bibitem{Tissier:2010ts}
M. Tissier and N. Wschebor,
\textit{Phys. Rev. D} \textbf{82} (2010) 101701
[arXiv:1004.1607 [hep-ph]].

\bibitem{Tissier:2011ey}
M. Tissier and N. Wschebor,
\textit{Phys. Rev. D} \textbf{84} (2011) 045018
[arXiv:1105.2475 [hep-th]].

\bibitem{Pennington:2011wd}
M. Pennington and D. Wilson,
arXiv:1109.2117 [hep-ph].

\bibitem{Ferreira:2025anh}
M.~N.~Ferreira and J.~Papavassiliou,
Prog. Part. Nucl. Phys. \textbf{144}, 104186 (2025)
doi:10.1016/j.ppnp.2025.104186
[arXiv:2501.01080 [hep-ph]].

\bibitem{Boucaud:2011eh}
P.~Boucaud, D.~Dudal, J.~P.~Leroy, O.~Pene and J.~Rodriguez-Quintero,
JHEP \textbf{12}, 018 (2011)
doi:10.1007/JHEP12(2011)018
[arXiv:1109.3803 [hep-ph]].

\bibitem{Dudal:2012zx}
D.~Dudal, O.~Oliveira and J.~Rodriguez-Quintero,
Phys. Rev. D \textbf{86}, 105005 (2012)
doi:10.1103/PhysRevD.86.105005
[arXiv:1207.5118 [hep-ph]].

\bibitem{Huber:2025kwy}
M.~Q.~Huber, C.~S.~Fischer and H.~Sanchis-Alepuz,
Eur. Phys. J. C \textbf{85}, no.8, 859 (2025)
doi:10.1140/epjc/s10052-025-14590-3
[arXiv:2503.03821 [hep-ph]].

\bibitem{Sternbeck:2009yx}
A. Sternbeck, E. Ilgenfritz, M. Muller-Preussker and A. Schiller,
PoS LAT2009 (2009) 210
[arXiv:1003.1585 [hep-lat]].

\bibitem{Boucaud:2006xe}
P. Boucaud, J. Leroy, A. Le Yaouanc, J. Micheli, O. Pene and J. Rodriguez-Quintero,
\textit{Phys. Rev. D} \textbf{74} (2006) 034505
[hep-ph/0507104].

\bibitem{Boucaud:2009ga}
P. Boucaud, F. De Soto, J. Leroy, A. Le Yaouanc, J. Micheli, O. Pene and J. Rodriguez-Quintero,
\textit{Phys. Rev. D} \textbf{79} (2009) 014508
[arXiv:0811.2059 [hep-ph]].


\bibitem{Taylor:1971ff}
J. Taylor,
\textit{Nucl. Phys. B} \textbf{33} (1971) 436.

\bibitem{Cucchieri:2004sq}
A.~Cucchieri, T.~Mendes and A.~Mihara,
JHEP \textbf{12}, 012 (2004)
doi:10.1088/1126-6708/2004/12/012
[arXiv:hep-lat/0408034 [hep-lat]].

\bibitem{Cucchieri:2006tf}
A.~Cucchieri, A.~Maas and T.~Mendes,
Phys. Rev. D \textbf{74}, 014503 (2006)
doi:10.1103/PhysRevD.74.014503
[arXiv:hep-lat/0605011 [hep-lat]].

\bibitem{Cucchieri:2008qm}
A.~Cucchieri, A.~Maas and T.~Mendes,
Phys. Rev. D \textbf{77}, 094510 (2008)
doi:10.1103/PhysRevD.77.094510
[arXiv:0803.1798 [hep-lat]].

\bibitem{Mintz:2017qri}
B.~W.~Mintz, L.~F.~Palhares, S.~P.~Sorella and A.~D.~Pereira,
Phys. Rev. D \textbf{97}, no.3, 034020 (2018)
doi:10.1103/PhysRevD.97.034020
[arXiv:1712.09633 [hep-th]].

\bibitem{Barrios:2024idr}
N.~Barrios, M.~Pel{\'a}ez, M.~S.~Guimaraes, B.~W.~Mintz and L.~F.~Palhares,
Phys. Rev. D \textbf{109}, no.9, 094039 (2024)
doi:10.1103/PhysRevD.109.094039
[arXiv:2402.17534 [hep-ph]].

\bibitem{Barrios:2020ubx}
N.~Barrios, M.~Pel{\'a}ez, U.~Reinosa and N.~Wschebor,
Phys. Rev. D \textbf{102}, 114016 (2020)
doi:10.1103/PhysRevD.102.114016
[arXiv:2009.00875 [hep-th]].

\bibitem{Figueroa:2021sjm}
F.~Figueroa and M.~Pel{\'a}ez,
Phys. Rev. D \textbf{105}, no.9, 094005 (2022)
doi:10.1103/PhysRevD.105.094005
[arXiv:2110.09561 [hep-th]].

\bibitem{Ilgenfritz:2006he}
E.~M.~Ilgenfritz, M.~Muller-Preussker, A.~Sternbeck, A.~Schiller and I.~L.~Bogolubsky,
Braz. J. Phys. \textbf{37}, 193-200 (2007)
doi:10.1590/S0103-97332007000200006
[arXiv:hep-lat/0609043 [hep-lat]].

\bibitem{Sternbeck:2006rd}
A.~Sternbeck,
[arXiv:hep-lat/0609016 [hep-lat]].

\bibitem{Collins:2016aya}
J.~C.~Collins and J.~A.~M.~Vermaseren,
[arXiv:1606.01177 [cs.OH]].

\bibitem{Aguilar:2024bwp}
A.~C.~Aguilar, N.~Brito, M.~N.~Ferreira, J.~Papavassiliou, O.~Oliveira and P.~J.~Silva,
Phys. Lett. B \textbf{858}, 139054 (2024)
doi:10.1016/j.physletb.2024.139054
[arXiv:2404.06496 [hep-lat]].

\bibitem{Colaco:2024gmt}
M.~Cola{\c{c}}o, O.~Oliveira and P.~J.~Silva,
Phys. Rev. D \textbf{109}, no.7, 074502 (2024)
doi:10.1103/PhysRevD.109.074502
[arXiv:2401.12008 [hep-lat]].

\bibitem{Oliveira:2025tfq}
O.~Oliveira, M.~Cola{\c{c}}o and P.~J.~Silva,
PoS \textbf{LATTICE2024}, 391 (2025)
doi:10.22323/1.466.0391
[arXiv:2501.17650 [hep-lat]].

\bibitem{Oliveira:2025rzw}
O.~Oliveira and P.~J.~Silva,
Phys. Rev. D \textbf{113}, no.7, 074512 (2026)
doi:10.1103/vyqj-7x47
[arXiv:2512.19839 [hep-lat]].

\bibitem{Brito:2024aod}
N.~Brito, O.~Oliveira and P.~J.~Silva,
PoS \textbf{LATTICE2024}, 469 (2025)
doi:10.22323/1.466.0469
[arXiv:2411.17280 [hep-lat]].

\bibitem{Brito:2025kfz}
N.~Brito, M.~Cola{\c{c}}o, O.~Oliveira and P.~J.~Silva,
PoS \textbf{QCHSC24}, 267 (2025)
doi:10.22323/1.483.0267
[arXiv:2505.23476 [hep-lat]].

\bibitem{Chroma}
R. G. Edwards et al. [SciDAC, LHPC and UKQCD],
Nucl. Phys. B Proc. Suppl. 140, 832 (2005)
doi:10.1016/j.nuclphysbps.2004.11.254 [arXiv:hep-lat/0409003 [hep-lat]].

\bibitem{PFFT}
M. Pippig, SIAM J. Sci. Comput. 35, C213 (2013).

\bibitem{Davies:1987vs}
C.~T.~H.~Davies, G.~G.~Batrouni, G.~R.~Katz, A.~S.~Kronfeld, G.~P.~Lepage, K.~G.~Wilson, P.~Rossi and B.~Svetitsky,
Phys. Rev. D \textbf{37}, 1581 (1988)
doi:10.1103/PhysRevD.37.1581

\bibitem{Silva:2004bv}
P.~J.~Silva and O.~Oliveira,
Nucl. Phys. B \textbf{690}, 177-198 (2004)
doi:10.1016/j.nuclphysb.2004.04.020
[arXiv:hep-lat/0403026 [hep-lat]].

\bibitem{Leinweber:1998im}
D.~B.~Leinweber \textit{et al.} [UKQCD],
Phys. Rev. D \textbf{58}, 031501 (1998)
doi:10.1103/PhysRevD.58.031501
[arXiv:hep-lat/9803015 [hep-lat]].

\bibitem{Leinweber:1998uu}
D.~B.~Leinweber \textit{et al.} [UKQCD],
Phys. Rev. D \textbf{60}, 094507 (1999)
[erratum: Phys. Rev. D \textbf{61}, 079901 (2000)]
doi:10.1103/PhysRevD.60.094507
[arXiv:hep-lat/9811027 [hep-lat]].

\bibitem{Becirevic:1999uc}
D.~Becirevic, P.~Boucaud, J.~P.~Leroy, J.~Micheli, O.~Pene, J.~Rodriguez-Quintero and C.~Roiesnel,
Phys. Rev. D \textbf{60}, 094509 (1999)
doi:10.1103/PhysRevD.60.094509
[arXiv:hep-ph/9903364 [hep-ph]].

\bibitem{deSoto:2007ht}
F.~de Soto and C.~Roiesnel,
JHEP \textbf{09}, 007 (2007)
doi:10.1088/1126-6708/2007/09/007
[arXiv:0705.3523 [hep-lat]].

\bibitem{Vujinovic:2018nqc}
M.~Vujinovic and T.~Mendes,
Phys. Rev. D \textbf{99}, no.3, 034501 (2019)
doi:10.1103/PhysRevD.99.034501
[arXiv:1807.03673 [hep-lat]].

\bibitem{Catumba:2021hcx}
G.~T.~R.~Catumba, O.~Oliveira and P.~J.~Silva,
Phys. Rev. D \textbf{103}, no.7, 074501 (2021)
doi:10.1103/PhysRevD.103.074501
[arXiv:2101.04978 [hep-lat]].

\bibitem{Oliveira:2012eh}
O.~Oliveira and P.~J.~Silva,
Phys. Rev. D \textbf{86}, 114513 (2012)
doi:10.1103/PhysRevD.86.114513
[arXiv:1207.3029 [hep-lat]].

\bibitem{Suman:1995zg}
H.~Suman and K.~Schilling,
Phys. Lett. B \textbf{373}, 314-318 (1996)
doi:10.1016/0370-2693(96)00162-1
[arXiv:hep-lat/9512003 [hep-lat]].

\bibitem{Aguilar:2010cn}
A.~C.~Aguilar and J.~Papavassiliou,
Phys. Rev. D \textbf{83}, 014013 (2011)
doi:10.1103/PhysRevD.83.014013
[arXiv:1010.5815 [hep-ph]].

\bibitem{Rojas:2013tza}
E.~Rojas, J.~P.~B.~C.~de Melo, B.~El-Bennich, O.~Oliveira and T.~Frederico,
JHEP \textbf{10}, 193 (2013)
doi:10.1007/JHEP10(2013)193
[arXiv:1306.3022 [hep-ph]].

\bibitem{Oliveira:2020yac}
O.~Oliveira, T.~Frederico and W.~de Paula,
Eur. Phys. J. C \textbf{80}, no.5, 484 (2020)
doi:10.1140/epjc/s10052-020-8037-0
[arXiv:2006.04982 [hep-ph]].





\end{thebibliography}
\end{document}